\def   \ni {\noindent}

\def   \ssk {\vskip  5truept}

\def   \bsk {\vskip 15truept}
 
\def   \newpage {\vfill\eject}
\def   \newline {\hfil\break}

\documentstyle[epsfig]{article}
\begin{document}

\hsize 5truein
\vsize 8truein
\font\abstract=cmr8
\font\keywords=cmr8
\font\caption=cmr8
\font\references=cmr8
\font\text=cmr10
\font\affiliation=cmssi10
\font\author=cmss10
\font\mc=cmss8
\font\title=cmssbx10 scaled\magstep2
\font\alcit=cmti7 scaled\magstephalf
\font\alcin=cmr6 
\font\ita=cmti8
\font\mma=cmr8
\def\ref{\par\noindent\hangindent 15pt}
\null
 

\title{\ni NEAR-INFRARED PHOTOMETRY OF BLAZARS}

\bsk \bsk
\author{\ni C. Chapuis$^{1,2}$,
 S. Corbel$^{1}$,
 P. Durouchoux$^{1}$,
 T. N. Gautier$^{3}$, and
 W. Mahoney$^{3}$}
\bsk
\affiliation{1) Service d'Astrophysique DAPNIA, CEA Saclay
F-91191 Gif sur Yvette cedex
2) D\'epartement de physique, Universit\'e de Versailles,
F-78035 Versailles cedex
3) Jet Propulsion Laboratory 169-327, 4800 Oak Grove Dr., Pasadena, CA 91109
}                                                
\bsk
\baselineskip = 12pt

\abstract{ABSTRACT \ni

The rapid variability of blazars in almost
all wavelengths is now well established. Two days of observations were conducted
at the Palomar Observatory during the nights of 25 and 26 February 1997 with the
5-meter Hale telescope, in order to search for rapid variability in the near-infrared
(NIR) bands J, H, K$_{s}$ for a selection of eight blazars.
With the possible exception of 1156+295 (4C 29.45), no intraday or day-to-day
variability was observed during these two nights.  However, for these
eight blazars, we have measured the NIR
$\nu$F$_{\nu}$ luminosities and spectral indices.
 It has recently been reported that the $\gamma$-ray emission is better correlated
with the near-infrared luminosity than with the X-ray luminosity
 (Xie et al. 1997). This correlation is suggested as a general property of
 blazars because hot dust is the  main source of soft photons which
 are scattered off the relativistic jets of electrons to produce the
 gamma rays by inverse Compton scattering.  We thus used this relationship
to estimate the $\gamma$-ray luminosity.
}

\bsk
\baselineskip = 12pt
\bsk
\baselineskip = 12pt
 

\text{\ni 1. INTRODUCTION
\ssk
\ni

\bsk
\ni 
1.1 The blazar properties
\ssk
\ni
The discovery that blazars (i.e., optically violently variable 
quasars and BL Lac objects) and flat radio-spectrum quasars emit 
most of their power in high-energy gamma rays (Fichtel, et al. 1994) 
probably represents one of the most surprising results from the Compton Gamma-Ray 
Observatory (CGRO). Their luminosity above 100 MeV in some cases exceeds
10$^{48}$ ergs s$^{-1}$ (assuming isotropic emission) and can
 be larger 
(by a factor of 10-100) than the luminosity in the rest of the 
electromagnetic spectrum. Moreover, the $\gamma$-ray emission can be strongly 
variable on time-scales as short as days, indicating that the emission 
region is extremely compact (Kniffen et al. 1993). Blazars 
have smooth, rapidly variable, polarized continuum emission from radio 
through UV/X-ray wavelengths. All have compact flat-spectrum radio cores
and many exhibit superluminal motions.
\newpage
\bsk
\ni 
1.2 The origin of gamma rays in blazars
\ssk
\ni
A variety of theoretical models have been recently proposed to explain
the origin of the $\gamma$-ray emission of blazars.
Most models describing the high-energy emission involve 
beaming from a jet of highly relativistic particles and include:

(1) synchrotron self-Compton. The $\gamma$-ray spectrum is the high-energy 
extension of the inverse-Compton radiation responsible of the X-ray 
radiation (Maraschi et al. 1992), i.e., the scattering of synchrotron 
radiation by relativistic electrons gives rise to a higher frequency flux, 
which can be scattered a second time and so on.

(2) inverse Compton scattering of accretion-disk photons by relativistic 
nonthermal electrons in the jet (Dermer et al. 1992).

(3) inverse Compton scattering of ambient soft X-rays by relativistic 
pairs accelerated {\it in situ} by shock fronts in a relativistic jet (Blandford 
\& Levinson 1995).

(4) synchrotron emission by ultrarelativistic electrons and positrons 
(Ghisellini et al. 1993).

\ni
Various relations between the emission at different wavelengths are 
implied by these models and can be used to observationally distinguish among a 
variety of emission mechanisms.

\bsk
\ni
1.3. The infrared and near-infrared luminosities
\ssk
\ni
A strong correlation between $\gamma$-ray and near-infrared luminosities was 
recently reported for a sample of blazars and it was 
suggested that this relation might be a common property of these objects 
(Xie et al. 1997). For that reason, the authors conclude that hot dust is likely 
to be the main source of the soft photons (near-infrared) which are 
continuously injected within the knot and then produce $\gamma$-ray flares 
by inverse Compton scattering on relativistic electrons.
Given this correlation, it 
is easy to use the near-infrared luminosities to deduce the $\gamma$-ray fluxes,
and then, the total emitted fluxes.

\bsk
\ni 2. OBSERVATIONS
\ssk
\ni
We observed eight blazars
with the 5-meter Hale telescope on Mt. Palomar during the nights of 25 and 26 
February 1997, using the Cassegrain Infrared Camera, an
instrument based on a $256 \times 256$-pixel
InSb array with the J (1.25 $\mu$m), H (1.65 $\mu$m) and K$_{s}$ 
(2.15$\mu$m) filters and a field-of-view of 32 arcsec. 

The reduction of data was done under IRAF and included subtraction of 
the dark noise, flat field corrections, and combination of images to
remove bad pixels, cosmic rays, and the sky.  Then aperture photometry for each
object was performed using nearby faint standards for calibration.
The apparent magnitudes are summarized in Table 1 and plotted in Figure 1. 

\begin{figure}[t]
\centerline{\psfig{file=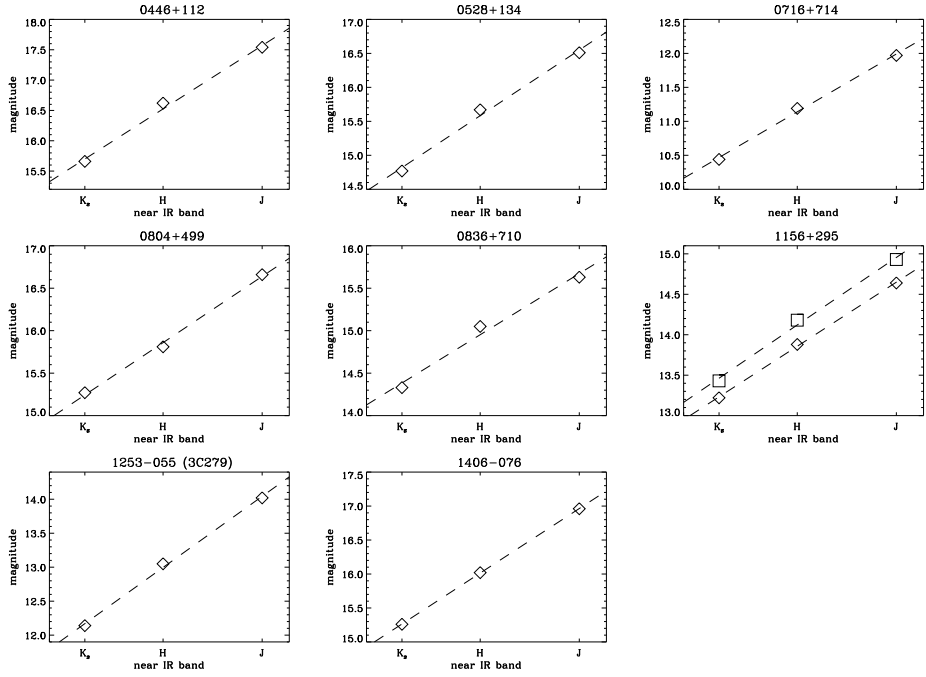,width=12.5cm,clips=,bbllx=50pt,bblly=360pt,bburx=550pt,bbury=730pt}}
\caption{FIGURE 1. The mean magnitudes of steady blazars are drawn with diamonds 
in J, H and K$_{s}$ bands.  For 1156+295 squares (25 February) and diamonds
(26 February) are used to show the variation between the two nights.
Typical errors range from 0.05 to 0.08 magnitudes.  The power law fit is indicated by
the dashed lines.
}
\end{figure}

 Due to the steadyness of the sources, it was possible to fit the energy flux
to a power law (defined as $f(\nu) \propto {\nu}^{-\alpha}$) by
$\chi^{2}$ minimization, giving the spectral index, $\alpha$, for each source (Table 1).
Finally, we calculated the luminosity, 
$L(\nu) = 4\pi{d_{L}^2} \nu f(\nu)$,

(1) using the luminosity distance, $d_L$, where 
$q_{0}=0.5$, $H_{0}=50 $ km s$^{-1}$ Mpc$^{-1}$,
z is the redshift, and c the velocity of light in vacuum:
\begin{center}
	$d_{L}=\frac{c}{H_{0}q_{0}^{2}}(zq_{0}-(1 - q_{0})(\sqrt{1+2q_{0}z} - 1))$
	(Weinberg 1972)
\end{center}

(2) and the K-correction, where $\alpha$ is the spectral index:

  \begin{center}
 $f(\nu)=f_{obs}(\nu) (1+z)^{(\alpha-1)}$.
\end{center}

\noindent  The K-corrected
   $\overline{L}_{\nu}$ luminosities in the K$_{s}$-band (as defined in Dondi \& 
   Ghisellini (1995)), calculated from our near-infrared observations, are given in
 Table 1.  For 0716+714, we took the lower limit $z > 0.3$ of Wagner et al.
 (1996). All other redshifts were taken from
   compilations of Ghisellini et al. (1993) and Dondi \& Ghisellini (1995). 

A strong correlation observed between $\gamma$-ray and near-infrared
 luminosities was shown (Xie et al. 1997) and the authors suggest that it may be a
  common property of blazars. According to them, inverse Compton 
  scattering of the infrared radiation from hot circumnuclear dust
 by a relativistic electron beam should be responsible for the $\gamma$-ray flares. 
According to Xie, et al. (1997), the near-infrared and $\gamma$-ray luminosities of
blazars can be related by:

\begin{center}

$\log{\overline{L}_\gamma} = 1.26 \log{\overline{L}_{IR}} - 11.38 $

\end{center}
Using this relationship, we then estimated the $\nu$F$_{\nu}$ luminosity in the  $\gamma$-ray range, using the K-corrected
$\overline{L}_{\nu}$ luminosity in the $K_{s}$-band.
These results are summarized in Table 1 and their
 discussion is given in Chapuis et al. (1998).
  
\begin{table}
\begin{center}\scriptsize
\begin{tabular}{|c|c|ccc|c|r|r|}
\hline
 & & & & & & & \\
IAU Name & $z$ & \multicolumn{3}{c|}{Apparent Magnitude}   & $\alpha$ &  $\overline{L}_{IR}$  & $\overline{L}_{\gamma}$ \\
    & & J & H & K$_s$ & & & \\
\hline
\hline
0446+112 &  1.207 & 17.54(10) &  16.62(6) &  15.66(5)  & 1.6 & 7.3 & 25\\
\hline
0528+112 &  2.060 & 16.51(5)  &  15.67(5) &  14.77(5)  & 1.4 &  53 & 311\\
\hline
0716+714 & $>0.3$ & 11.97(5)  &  11.19(5) &  10.44(5)  & 1.0 & $>27$ & $>134$ \\
\hline
0804+499 &  1.43  & 16.66(5)  &  15.81(5) &  15.27(5)  & 0.6 &  9.4 & 35\\
\hline
0836+710 &  2.172 & 15.63(5)  &  15.05(5) &  14.33(7)  & 0.6 &  36 & 191\\
\hline
1156+295 &  0.729 & 14.93(5)  &  14.18(5) &  13.43(5)  & 0.9 &  11 & 43\\
         &        & 14.64(5)  &  13.88(5) &  13.22(5)  & 0.8 &  12.6 & 51\\
\hline
1253-055 (3C279)&0.538&14.02(3)& 13.05(5) &  12.14(5)  & 1.5 &  25 & 124\\
\hline
1406-076 &  1.494 & 16.96(5)  &  16.02(5) &  15.26(5)  & 1.2 &  14 & 57\\
\hline
\end{tabular}
\end{center}
\caption{Table 1. Summary of blazar observations where the columns represent
(1) IAU name, (2) redshift, (3-5) J, H, and K$_s$ magnitudes, (6) near-infrared
spectral index, and (7-8) K-corrected near-infrared
($\overline{L}_{IR}$) and gamma-ray ($\overline{L}_{\gamma}$) luminosities in
units of $10^{45}$ erg s$^{-1}$.}
\end{table}

{\bf Acknowledgement} Observations at the Palomar Observatory were made as part of a
continuing collaborative agreement between Palomar Observatory and the
Jet Propulsion Laboratory.  The research described in this paper was carried out
in part by the Jet Propulsion Laboratory, California Institute of Technology,
under contract to the National Aeronautics and Space Administration.
}

\bsk
\baselineskip = 12pt
 
 
{\references \ni REFERENCES
\ssk

\ref Blandford, R.D., \& Levinson, A. 1995, ApJ, 441, 79

\ref Chapuis, C., et al. 1998 (in preparation)

\ref Dermer, C., Schlickheiser, R., \& Mastichiadis, A. 1992, A\&A, 256, L27

\ref Dondi, L., \& Ghisellini, G. 1995, MNRAS, 273, 583

\ref Fichtel, C. E., et al. 1994, ApJS, 94, 551

\ref Ghisellini, G., Padovani, P., Celotti, A., \& Maraschi, L. 1993, ApJ, 407, 65

\ref Kniffen, D. A., et al. 1993, ApJ, 411, 133

\ref Maraschi, L., Ghisellini, G., \& Celotti, A. 1992, ApJ, 397, L5

\ref Wagner, S.J., et al. 1996, AJ, 111, 2187

\ref Weinberg, S. 1972, Gravitation and Cosmology, John Wiley \& Sons NY

\ref Xie, G., Zhang, Y., \& Fan, J. 1997, ApJ, 477, 114
}
\end{document}